\documentclass[widetext,showpacs,preprintnumbers,amsmath,amssymb]{revtex4}

\usepackage{graphicx}
\usepackage{dcolumn}
\usepackage{bm}
\begin{document}


\title{Network properties of written human language}
\author{A. P. Masucci}
\author{G. J. Rodgers}%
\affiliation{%
Department of Mathematical Sciences, Brunel University,
Uxbridge, Middlesex, UB8 3PH, United Kingdom}%

\date{\today}
\begin{abstract}
 We investigate the nature of written human language
within the framework of complex network theory.   In particular,
we analyse
 the topology of  Orwell's \textit{1984} focusing on the
local properties of the network, such as the properties of the
nearest neighbors and the clustering coefficient. We find a
composite power law behavior for both the  average nearest
neighbor's degree and average clustering coefficient as a function
of the vertex degree. This implies the existence of different
functional classes of vertices. Furthermore we find that the second
order vertex correlations are an essential component of the network
architecture. To model our empirical results we extend a previously
introduced model for language due to Dorogovtsev and Mendes.   We
propose an accelerated growing network model that contains three
 growth mechanisms: linear preferential attachment, local
preferential attachment and the random growth of a pre-determined
small finite subset of  initial vertices.
 We find that with these elementary stochastic rules we are able to
 produce a network showing syntactic-like structures.
\end{abstract}
 \maketitle

\section{\label{sec:level1}Introduction.\protect}
Many systems in nature are composed of a large number of interacting
agents that exhibit small world, scale free and hierarchical
behavior. Such networks can be found in several disciplines
including social human organization, biological and chemical
structures, the world wide web, etc...\cite{5}.

 Words are a good example of simple
elements that  combine to form complex structures such as novels,
poems, dictionaries and manuals that are designed to transport or
convey information. The written human language is one of the most
important examples of a self-organizing system in nature. It is not
only interesting in a linguistic and philosophical sense, but it
also has the virtue of being a accessible network system which
allows full empirical studies of its structure to be carried out. If
we consider a complete single book as our system and treat this book
as a finite directed network in which the words are the vertices and
two vertices are linked if they are neighbors, then we can analyse
this network completely. We are able to know everything about the
construction of such a network: we know the sequence in which each
vertex was linked to the network and we can follow the growth of
each vertex degree without error. Moreover, we already know a great
deal about the syntactic and logical structure of grammar which
determines how the network organises itself. Finally, plenty of data
is available, covering centuries of human literature.

The first quantitative observation that language displays a
complex topological structure was due to  Zipf in 1949 \cite{1}.
Zipf observed that if the words of a prose are ordered in their
rank of occurrence, that is the most frequent word in the prose
has rank one, the second rank two and so on, and  the frequency of
the words is plotted against their rank, then, for every text and
for every language, we find a  skewed distribution that fits a
power law (see Fig.\ref{f2}a as an example). This universal
empirical law is known as "Zipf's Law".

The first attempt to explain this law was due to Simon in 1955
\cite{3}. He proposed a stochastic model for generating a text based
on the frequency of occurrence of words, which he was able to solve
exactly. In this model a stochastic text was built by adding a
previously used word with probability  proportional to its frequency
and by adding  new words with constant rate.

With the introduction of scale free network theory \cite{4, 5},
human language structure was examined by a number of authors
\cite{8, 9, 10, 11, 12}. There is a straight forward connection
between  the rank $r$ of a word and the scale free distribution for
the vertices degree in a network.  If we define the degree $k$ of a
word as the number of different words this word is connected to, and
$P(k)$ the word degree distribution, we have:
\begin{equation}\label{0}
  r(k)\propto\int_{k}^{\infty}P(k')dk'.
\end{equation}
 In language network the degree of a
word is equivalent to its frequency, so that Eq. (\ref{0}) is a
direct link to transform the scale free  degree distribution into
Zipf's Law.
 Thus, in the context of growing
networks, Simon's model is equivalent to the more recent \cite{dor},
where the network growth is regulated by preferential attachment
\cite{4}.

In all  the  models above, growth is based on the global properties
of the text. These classical models display a good power law for the
degree distribution, but, as it will be stressed later, this power
law distribution holds even when we randomize the words in the text,
that is if we write a meaningless book. Thus,  the degree
distribution is not the best measure of the self organizing nature
of this network.

 Nevertheless, everybody who is
experienced with the process of writing knows that syntax is the
basic rule used to build a sentence. Syntax is nothing more than a
set of local rules that give the ensemble of words in a phrase an
intelligent and understandable structure.

In this work we analyse  in detail the topology of the written
language network focusing on the local properties and we find that
these local properties are
 essential elements of the network architecture. We find
that to build a stochastic model reproducing the main properties of
language we need several growth mechanisms for the network. We
obtain the best fits with real data by adding a random attachment
mechanism, for a small preselected set of vertices, in an
accelerated growing network, that displays both global and local
preferential attachment.

The paper is organized as follow: in section II we  analyse the
growth properties and the nearest neighbors properties of a novel,
in section III we generalize the Dorogovtsev-Mendes model for
language \cite{10} and  extend it to include the local behavior of
vertices.

\section{Topology of the Network}
We consider the George Orwell's   \textit{1984} as our system
\cite{orw, 19}. This novel can be treated as a finite directed
network in which the words are the vertices and two vertices are
linked if they are neighbors. Punctuation is also considered as
vertices.

 For our purposes it is very important to consider written
human language as a directed network since the syntactic rules of
language are not reflexive. An interesting measure to quantify this
behavior of language is the link reciprocity, that quantifies the
non-random presence of mutual edges between pairs of vertices
\cite{garl}. Reciprocity values lie between -1 for perfectly
antireciprocal networks, 0 for areciprocal networks and 1 for
perfectly reciprocal networks. In our case we find  a value of 0.021
for the novel \textit{1984}. It means that written human language is
one of the natural networks closest to the absolute areciprocity.

\begin{figure*}
\includegraphics[width=9cm]{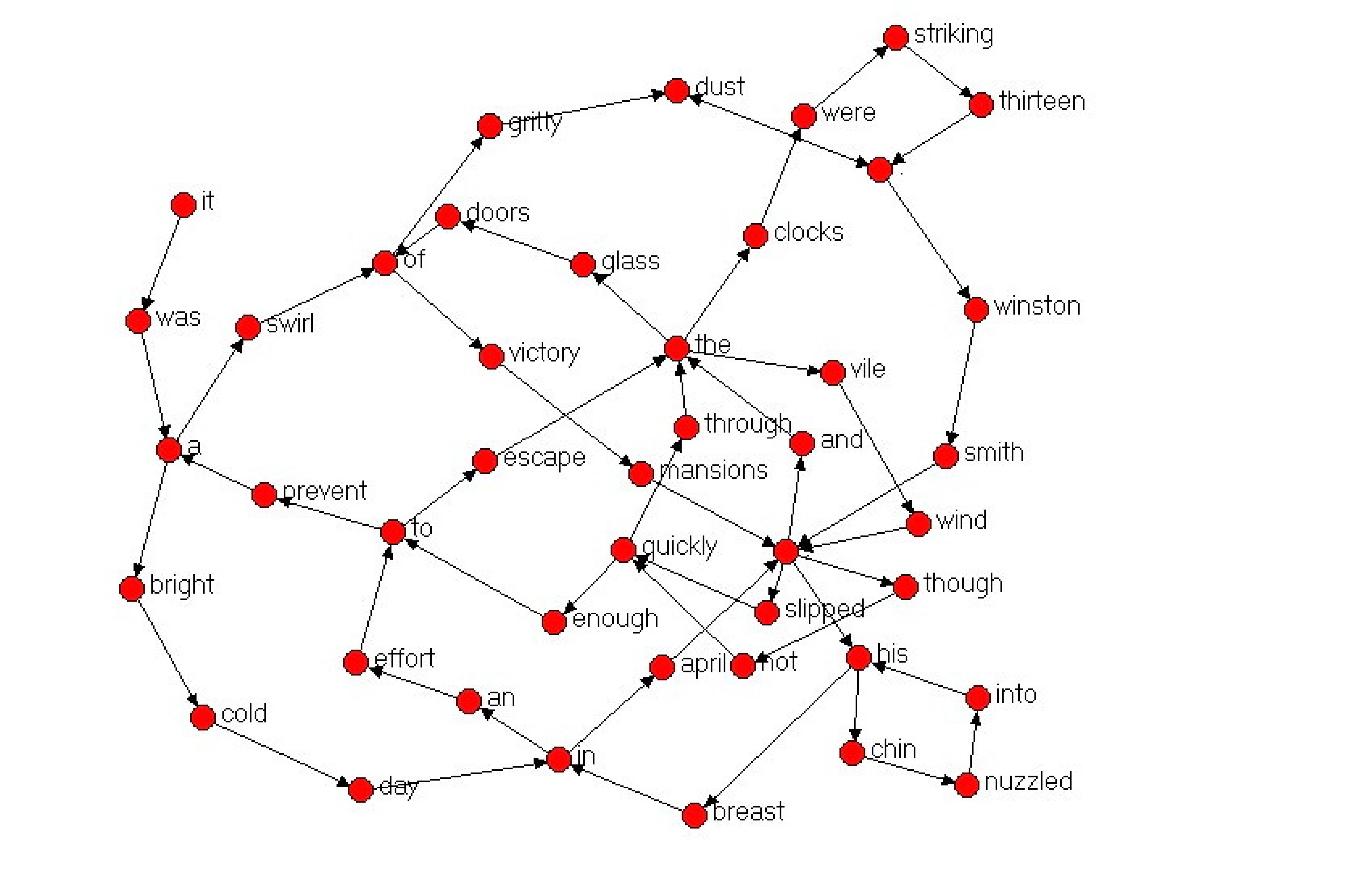}
\caption{\label{f1} Illustration of the language network for the
first 60 words of Orwell's \textit{1984}.}
\end{figure*}

In Fig.\ref{f1} we show the first 60 vertices of our network,
corresponding to the first 60 words of \textit{1984}. A random walk
in the whole network defines exactly all the grammatical and
syntactical rules that are in the text. In such a directed network
the degree and the frequency of a word have the same meaning, and
are equal, because every time a new word is added to the text it is
the only vertex of the network to acquire an edge.

\begin{figure}[!ht]\center
         \includegraphics[width=0.48\textwidth]{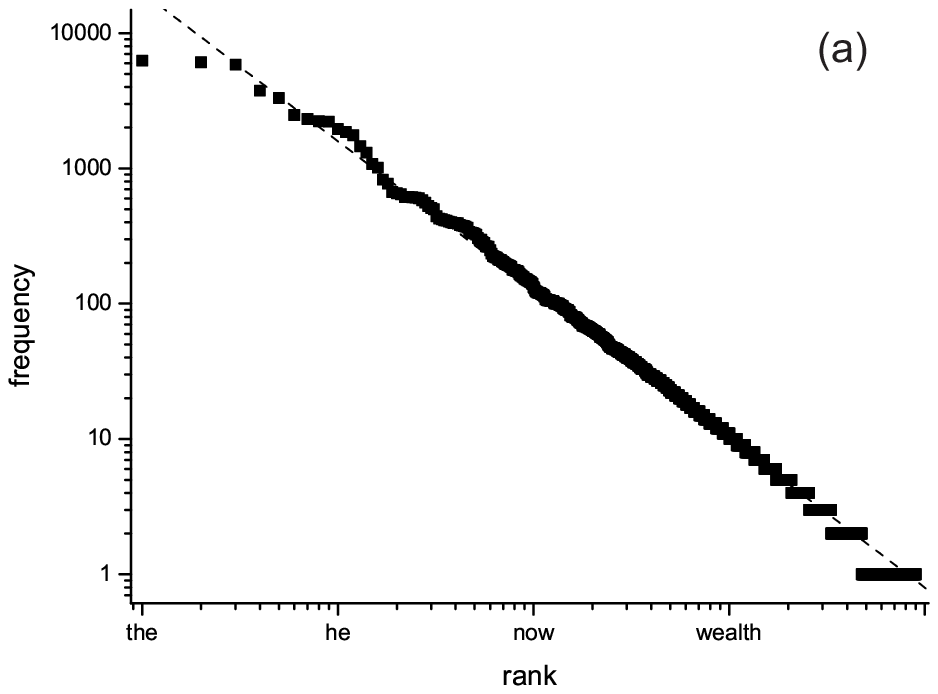}
         \includegraphics[width=0.48\textwidth]{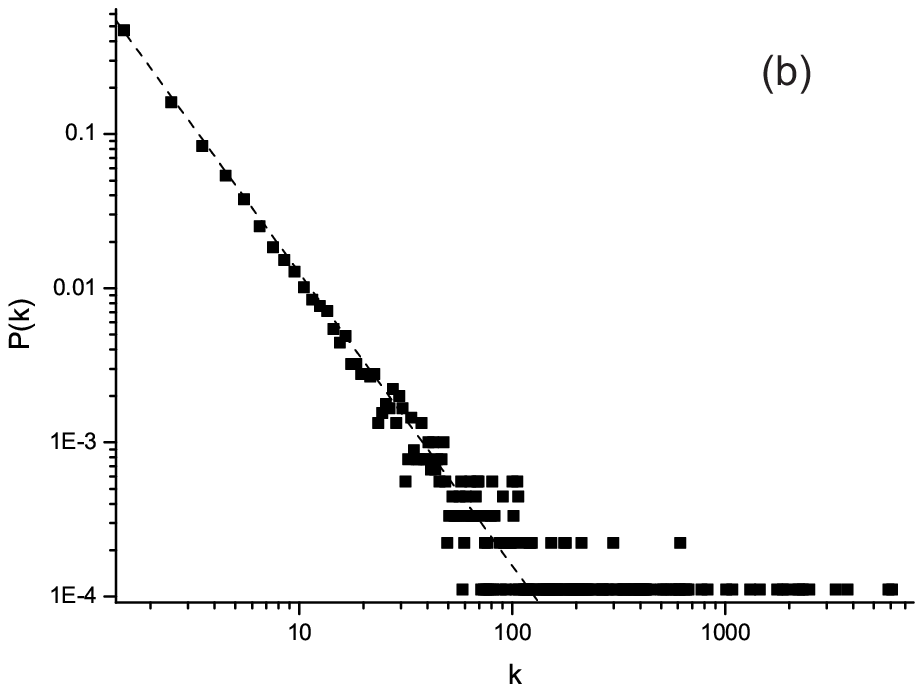}
         \caption{\label{f2} (a) Measure of the Zipf's Law on   \textit{1984}. The dashed
          line is a power law with slope -1.1. (b) The degree distribution P(k) measured on the same novel. The slope found is -1.9.}
    \end{figure}

 The network is composed by 8992 vertices representing the different words and
 117687 directed edges so that the mean degree is $<k>=13.1$.
The degree distribution follows a power law with slope: $P(k)\propto
k^{-1.9}$ , and the Zipf's Law, according to eq.(\ref{0}), has slope
-1.1(see Fig.\ref{f2}).

It is important to notice that the feature of being a scale free
network doesn't depend on the syntactic structure of the language
\cite{16}. In fact, due to the equivalence between frequency and
degree of a vertex, if we shuffle the text we obtain the same degree
distribution, but loose all the syntactic structure.

\subsection{Growth properties}

 A basic property of  language
network is that it is an accelerated growing network in the sense
that the number of edges between words grows faster than the number
of vertices \cite{14}. If we define the time step $t$ as the time in
which a new word is added to the text and $N(t)$ the total number of
words in the text at time $t$,  we find empirically  that
\begin{equation}\label{2}
  N(t)\sim t^{1.8}.
\end{equation}
At every time, $t$ will represent the number of different words used
to compose the text, or the used vocabulary size, while $N(t)$ will
represent the total novel size.

In order to build a stochastic model for language, we will need a
random growth mechanism that attaches new words to the text and a
preferential attachment (PA) mechanism that attaches previously used
words to the text. Moreover we will need a mechanism that could
catch the inner structure of the text, the syntax.

\subsection{Nearest neighbor's
properties}

Syntax is made up of local and selective structures that can be
recognizable through the analysis of nearest neighbors.
 A very important measure to quantify the hierarchical structure of a network is the clustering coefficient \cite{17}.
This counts the triangles that form in the unweighted and directed
network associated with the network in consideration. We expect
language to show a low average clustering coefficient because only a
few triangles are present in the related network. The reason for
this is the selectiveness of syntactic structures. For example the
word "like" is able to link to definite and indefinite articles "a"
and "the" but these articles will never be linked to each other.

 We define the clustering coefficient $c_{i}$ for every vertex $i$ of our network as
\begin{equation}\label{1}
  c_{i}\equiv\frac{e_{i}}{d_{i}(d_{i}-1)},
\end{equation}

where $d_{i}$ is the number of the different nearest neighbors of
vertex $i$ (with $d_{i}\neq$ 0, 1) and $e_{i}$ is the number of
directed edges that connect those nearest neighbors. This formula is
a generalization of that for undirected networks \cite{5}.

  We find  that the mean clustering coefficient for
our network is $<c>=0.19$, that is the clustering coefficient is on
average very low if compared to that of other real networks
\cite{5}. This  is due to the syntactic structure of  language that
tends to create functional structures instead of clustering
structures.

\begin{figure*}
\includegraphics[width=8.0cm]{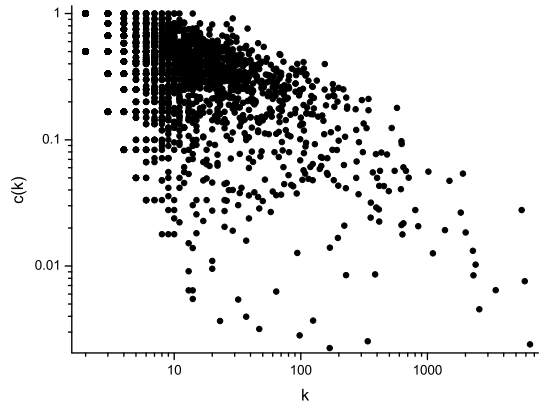}
\caption{\label{f3} The clustering coefficient against the degree
of the vertices in \textit{1984}.}
\end{figure*}

In Fig.\ref{f3} we show the  clustering coefficient against the
degree of the vertices for our novel. The clustering coefficient
values are  spread across the graph. If we associate those values to
the properties of the subgraph associated to each word, we
understand how words can display a very complex organization.

\begin{figure}[!ht]\center
         \includegraphics[width=0.48\textwidth]{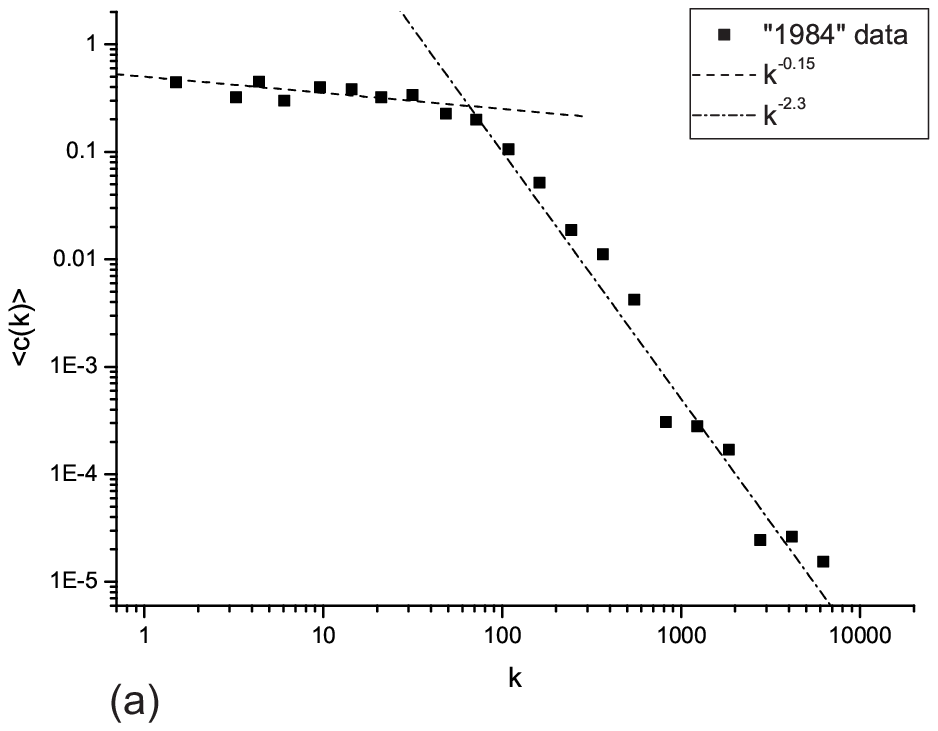}
         \includegraphics[width=0.48\textwidth]{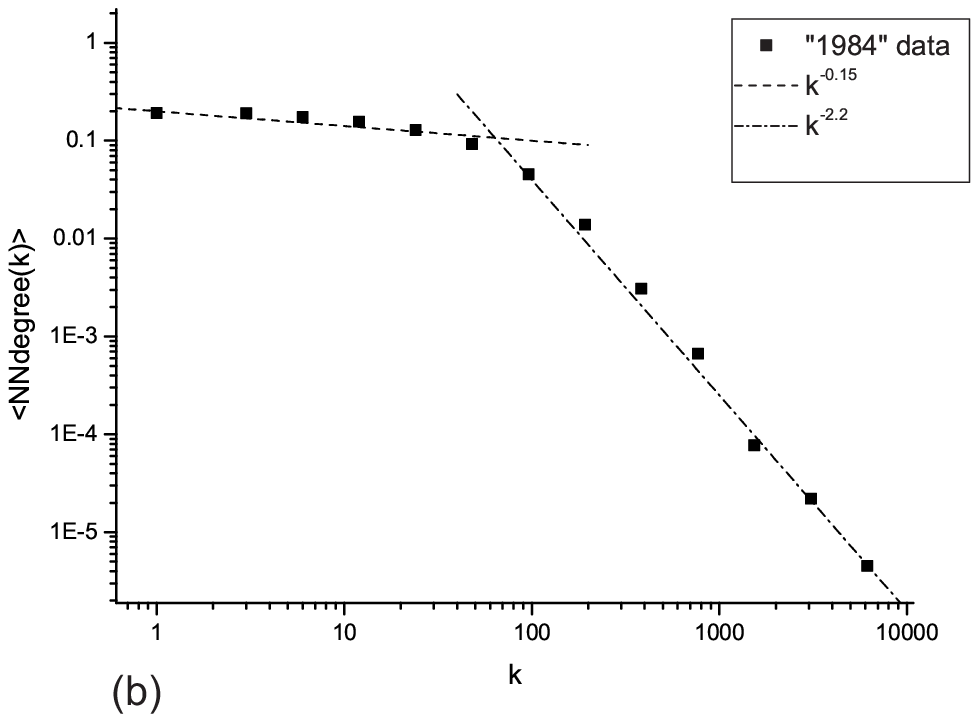}
         \caption{\label{f4} (a) The average clustering coefficient as a function
          of the vertex degree. (b) The average nearest neighbors degree as a function of the vertex degree.
          Noise has been removed using logarithmic binning. In both plots two different regions emerge displaying different power law behavior. }
    \end{figure}

 Even if the mean clustering coefficient for this kind of network is very low, modularity is still evident from a global measure of
 the mean clustering coefficient and the mean nearest neighbors degree as a function of the vertex degrees.
 In particular
 the mean clustering coefficient as a function of the vertex degree is flat for random graphs \cite{18}.
 Our measures are shown in Fig.\ref{f4}a.
Two different behaviors clearly emerge. For low values of $k$ the
data is nearly flat, which means low degree vertices don't display a
strong hierarchical behavior. It's not the same for high degree
vertices where hierarchical structures are apparent.

 The mean nearest neighbor's degree as a function of the vertex degree is a good estimator for the degree-degree correlations \cite{18}.
  This is also flat in the random graph case. Our measurements in  Fig.\ref{f4}b show the
   existence of global disassortative or negative correlations, where large degree vertices tend to be connected to those with low degree and vice versa.
Moreover we can see strong analogies with the measure of the mean
clustering coefficient. In fact the mean nearest neighbors degree
display two different behaviors, the first for low values of $k$
where the power law fitting curve is nearly flat, revealing very low
correlations between the degree of vertices. For large values of $k$
the power law behavior is much stronger, disclosing strong
degree-degree correlations for high degree vertices.

 The cut off in the  power laws for the mean clustering coefficient and the
  mean nearest neighbor's degree is nearly the same and  is around $k\approx100$.
  Only around 1\% of the total
number of vertices in the network posses $k>100$ and those
vertices belong to the 64\% of the total number of edges in the
network. Those vertices, belonging to the tale of $P(k)$
distribution (see Fig.\ref{f2}b), are essentially articles,
punctuation,
 prepositions and pronouns. They organize the main
 architecture of the network.

Many attempts to find a statistical measure that can describe the
syntactical structure as an emergent property of the language have
been proposed \cite{9, 15}, but these use a-priory information about
the logical role of each word.

The small value of the clustering coefficient implies, as we
stressed before, that there are only a few triangles in the network.
A particular structure emerging from the network is the directed
binary structure, that is when two words are linked together several
times in the text.

    \begin{figure}[!ht]\center
         \includegraphics[width=0.48\textwidth]{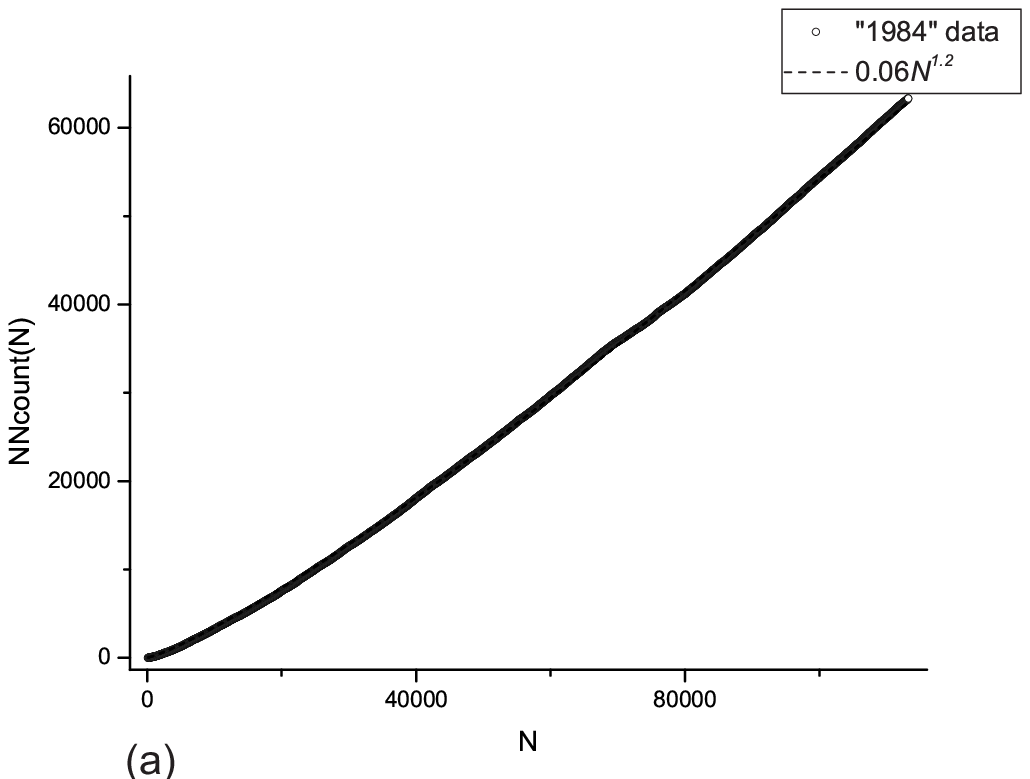}
         \includegraphics[width=0.48\textwidth]{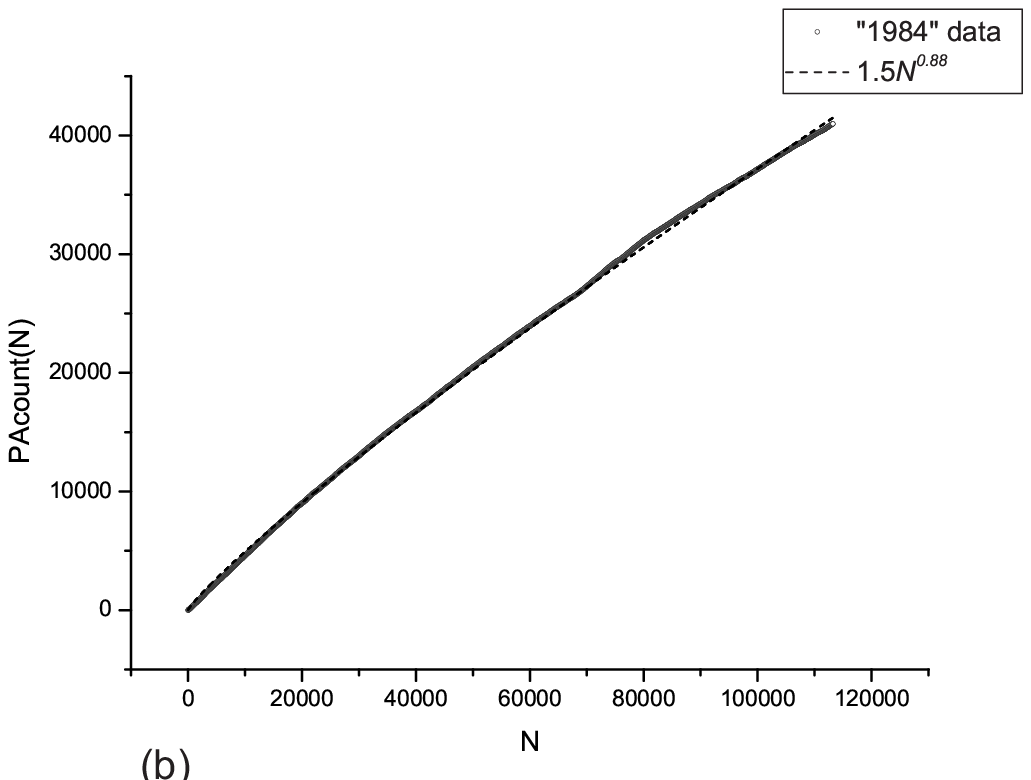}
         \caption{\label{f5} (a) Number of  occurrences of repeated binary structures of words
         in the text against the total number of words. (b) Number of  occurrences of previously existing words in the text, that are
           not part of previously existing binary structures, against the total number of words.}
    \end{figure}

\begin{table}[htbp]
\caption{\label{t1}The most frequent binary structures that are
present in \textit{1984} with their relative frequencies.   }
\begin{center}
\begin{tabular}{||c|c|c||}
\hline
Rank & Structure & Number of occurrences\\
\hline
1 & of the & 742\\
2 & , and & 717\\
3 & . the & 594\\
4 & it was & 576\\
5 & in the & 570\\
6 & . he & 560\\
7 & . it & 477\\
8 & ' s & 412\\
\hline
\end{tabular}
\end{center}
\end{table}

We show in Table \ref{t1} the most common binary structures we
find in our novel. Some of them can appear trivial, but looking at
the number of times they emerge in the text it is understandable
why we consider them so important. If we plot a histogram of the
occurrence of each binary structure we find a power law
distribution. This suggests that the binary structures play an
important role in the organization of the complex network.

In Fig.\ref{f5} we show a measure for the relative occurrence of
previously appeared words during the evolution of the text. In
Fig.\ref{f5}a we count the occurrence of a repeated word if it
belongs to a previously existing  binary structure, while in
Fig.\ref{f5}b we count the occurrence of a repeated word if it
doesn't belong to a previously existing binary structure. We find
that more than half the words added to the text  form binary
structures that are already present, while only 36\% of the words
entering the text doesn't belong to binary structures.

The standard way to create scale free networks is the PA mechanism
\cite{5}, that is a new vertex is linked to vertex $i$ with
probability proportional to its degree $k_{i}$. However, this
mechanism doesn't reproduce the massive formation of binary
structures observed in the language network.

Creation of binary structures implies that new edges form between
previously linked vertices. This behavior can be reproduced through
a stochastic process that is not the usual  PA, but  a
\textit{local} PA. By local PA we mean that a vertex will be linked
to a node \textit{i} chosen from its nearest neighborhood with
probability proportional to $k_{i}$.

It appears quite natural now, for the construction of a model, to
split the standard PA \cite{5} in the \textit{local} PA and the
\textit{global} PA. For global PA we mean a mechanism for choosing
vertices as the standard PA, but excluding the nearest neighborhood,
that is a vertex will be linked to a node $i$ that is not part of
its nearest neighborhood with probability proportional to $k_{i}$.

 In the next section we review an
important accelerated growing network model that fits very well with
global properties of  language and later we add to it the stochastic
behavior we found in the previous analysis to build up a model that
captures both the global and local properties of  real text.

\section{The Models}
Dorogovtsev and Mendes   introduced a model  \cite{10, 14}
(hereafter the \textit{D-M model}) for an accelerated growing
network as a basic model for language. We generalize this work with
the intention to understand the local properties of the language
such as the clustering coefficient and the nearest neighbor's
properties.

  We make a few modifications to the model to suit it to our analysis.
We will consider it as a directed network and, based on the
equivalence between frequency and degree of a word, we make some
simplifications in the implementation of the network.

  \subsection{D-M model}

   The D-M model \cite{10, 14} starts with a chain of 20 connected vertices.

1.  At each time step we add a new vertex (word) to the network. We
link it to an old vertex of the network through the standard PA,
that is it will be linked to any vertex $i$ with probability
proportional to $k_{i}$.

2. At its birth, $m(t)-1$ new edges emerge between old words, where
$m(t)$ is a linear function that can be measured and represents the
accelerated growth of the network in consideration. These new edges
emerge between old vertices $i$ and $j$ with the probability
proportional to the product of their degrees $k_{i}\cdot k_{j}$.

 We use
\begin{equation}\label{3}
  <m(t)>\approx0.002t+2.7 ,
\end{equation}
which is a result we measured from \textit{1984}.

\begin{figure}[!ht]\center
         \includegraphics[width=0.48\textwidth]{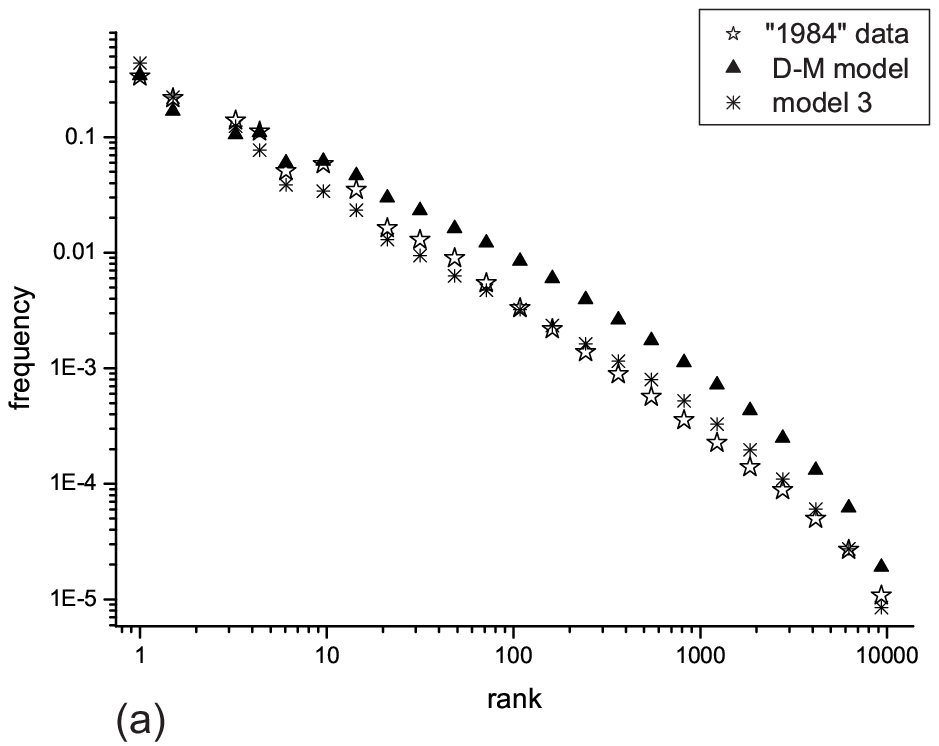}
         \includegraphics[width=0.48\textwidth]{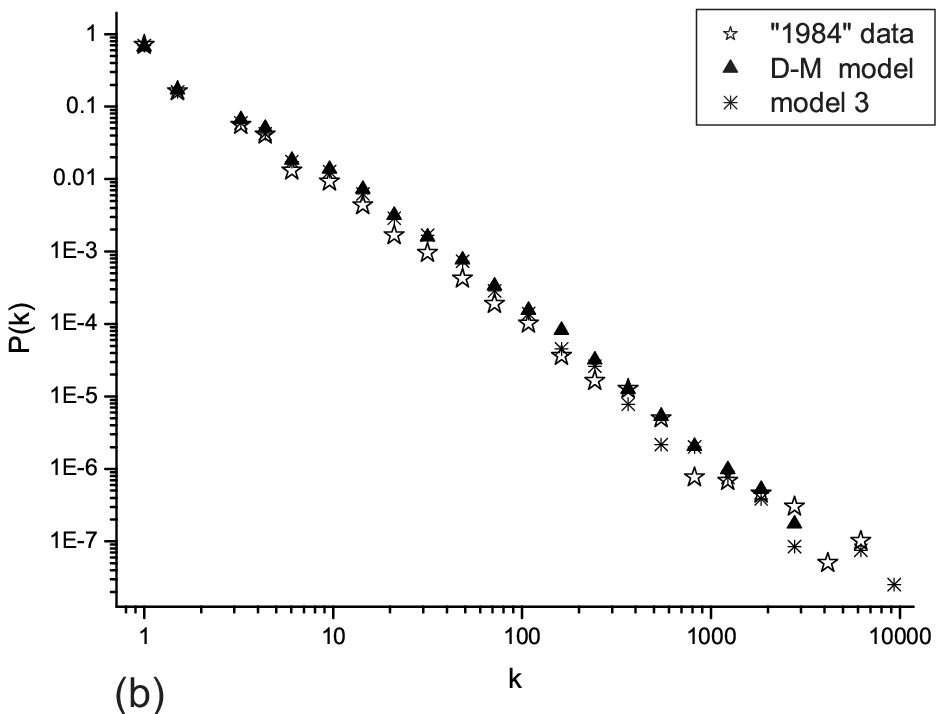}
         \caption{\label{f7} Comparison between the \textit{D-M model}, \textit{model 3} and real
          data. (a) Zipf's Law. (b) Degree frequency count. In both plots logarithmic binning is used to reduce noise. }
    \end{figure}

 This simple model has an analytical
solution and can reproduce very well the degree distribution
\cite{10, 14}. However it fails to reproduce  Zipf's Law (see
Fig.\ref{f7}a) and
 the internal structure of the language. The average
clustering coefficient measured in \textit{D-M model} is
$<c(k)>=0.16$, that is smaller then that measured for \textit{1984},
but close to it.

In Fig.\ref{f8}a we show the clustering analysis performed on the
model compared to our network. Even though  the general behavior
of the model follows that of the real data, big differences are
quite evident. $<c(k)>$ for the model is much narrower then  that
from the empirical measurements. This means that it doesn't catch
the main complex organization of words. In the text words belong
to different subgraphs, reflected in their clustering coefficient,
depending on their functional role. In the model all vertices look
equivalent in a single global hierarchical organization.

    \begin{figure}[!ht]\center
         \includegraphics[width=0.48\textwidth]{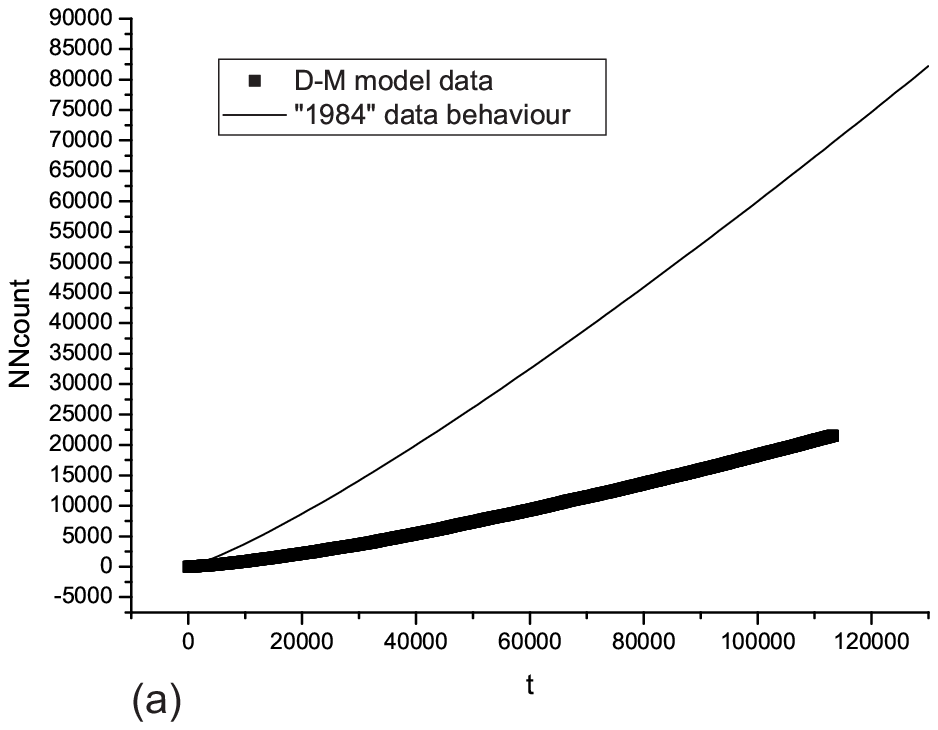}
         \includegraphics[width=0.48\textwidth]{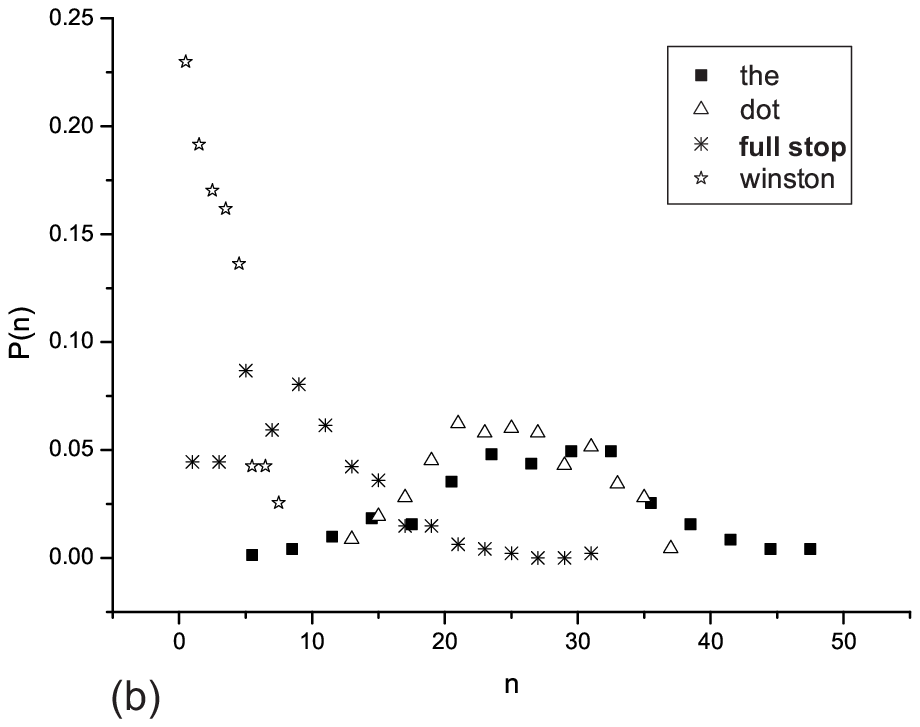}
         \caption{\label{f6} (a) Count of the occurrence of binary structures during the evolution
         of the network for the \textit{D-M model} compared to the real network data. (b) Probability
         distribution for the occurrence of some of the most frequent words in \textit{1984}. To obtain this measure we partitioned
         the novel, each partition of 500 words, and counted the number of times $n$ that  each word appears in each partition. As we can see different words display
         different distributions. In particular
"the" and the "full stop", that are structural words, display a
Gaussian distribution while a word as "Winston", who is the first
character in the novel, and is a meaningful word, follows an
extremely different distribution.}
    \end{figure}
Another measure we are interested in is the counting of the
occurrence of binary structures in the model during the evolution
of the network. We show this result in Fig.\ref{f6}a comparing it
with a line representing the measured value from our network. As
we can see the \textit{D-M model} misses the massive formation of
binary structures we  observe in the real network.

\subsection{ Extension of the model}

 We extend the D-M
model to include the local behavior of  language. We want to
elaborate the D-M approach distinguishing when  the PA attachment
mechanism is local, that is when a new word is attached to one of
its previous neighbors, or global.

We find that the probability that the preferentially chosen edges at
each time step are part of a previously existing binary structure
follows  the power law
\begin{equation}\label{4}
p(t)\approx 0.1t^{0.16} ,
\end{equation}

quite well. We try to implement this ingredient in the next model.

\textit{Model 2}: We start with a chain of 20 connected vertices.

1.  At each time step we add a new vertex (word) to the network.  We
link it to an old vertex of the network through the global PA, that
is it will be linked to the vertex $i$, that is not part of its
nearest neighborhood, with probability proportional to $k_{i}$.

2. For $m(t)-1$ times, where $m(t)$ is the measurable function
(\ref{3}), we perform the following operations: with probability
$p(t)$ we link the last linked vertex to an old vertex through local
PA, that is we link it to a node $i$ in its nearest neighborhood
with probability proportional to $k_{i}$; with probability $1-p(t)$
we link the last linked vertex to an old vertex through global PA,
that is the last linked vertex will be linked to the vertex $i$ that
is not part of its nearest neighborhood with probability
proportional to $k_{i}$ .

For this model the resulting average clustering coefficient is
very low if compared to that of our network. It means, as we could
expect, that the introduction of local PA supports selective local
rules in the growth of the network that strongly limits the
formation of triangles. We find $<c>=0.08$, while in \textit{1984}
$<c>=0.19$. Thus, by construction, \textit{model 2} matches the
analysis performed in Fig.\ref{f6}a for the growth of the network
due to repetition of binary structures.

\subsection{Model 3}

 The \textit{D-M model} catches the average clustering and the global
growth behavior of the network but misses the internal structure
while \textit{model 2} catches the global and nearest neighbor
growth behavior of the real network but not the characteristic
average clustering coefficient. Starting from this last model we
would like to find a mechanism to increase the value of the average
clustering coefficient.

It is now useful to consider the entropic analysis of language
performed by Zanette and Montemurro in \cite{17}. They found that
different words in written human language display different
statistical distributions, according to their function in the text.
Making a partition of the text and counting the occurrence of each
word in the partitions, they found that the most frequent words like
punctuation and articles follow  a Gaussian distribution, that is
they are randomly distributed.

We show in Fig.\ref{f6}b  a similar measure for the probability
distribution for the occurrence of different functional words in
the novel \textit{1984}. Our analysis agrees with that in
\cite{17}. With this in mind, we import into our  next model
three apriori selected vertices, representing main punctuation and
articles, with different growth properties to the other vertices
in the network.

\textit{Model 3}: We start with a chain of 20 linked vertices.

1. At each time step we add a new vertex (word) to the network.  We
link it to an old vertex of the network through the global PA, that
is it will be linked to the vertex $i$ that is not part of its
 nearest neighborhood with probability proportional to $k_{i}$.

2. For $m(t)-1$ times, where $m(t)$ is the measurable function
(\ref{3}), we perform the following operations:  with fixed
probability $q=0.05$ we link the last linked vertex to one of the
three fixed vertices; with probability $p(t)$, given by (\ref{4}),
we link the last linked vertex to an old vertex through local PA,
that is we link it to a node $i$ in its nearest neighborhood with
probability proportional to $k_{i}$; with probability $1-p(t)-3q$ we
link the last linked vertex to an old vertex through global PA, that
is the last linked vertex will be linked to the vertex $i$ that is
not part of its nearest neighborhood with probability proportional
to $k_{i}$.

In the last model, with the introduction of the new random
attachment mechanism, the average clustering coefficient increases
and becomes $<c>=0.20$, whilst preserving the global and the nearest
neighbors growth counting.

    \begin{figure}[!ht]\center
         \includegraphics[width=0.48\textwidth]{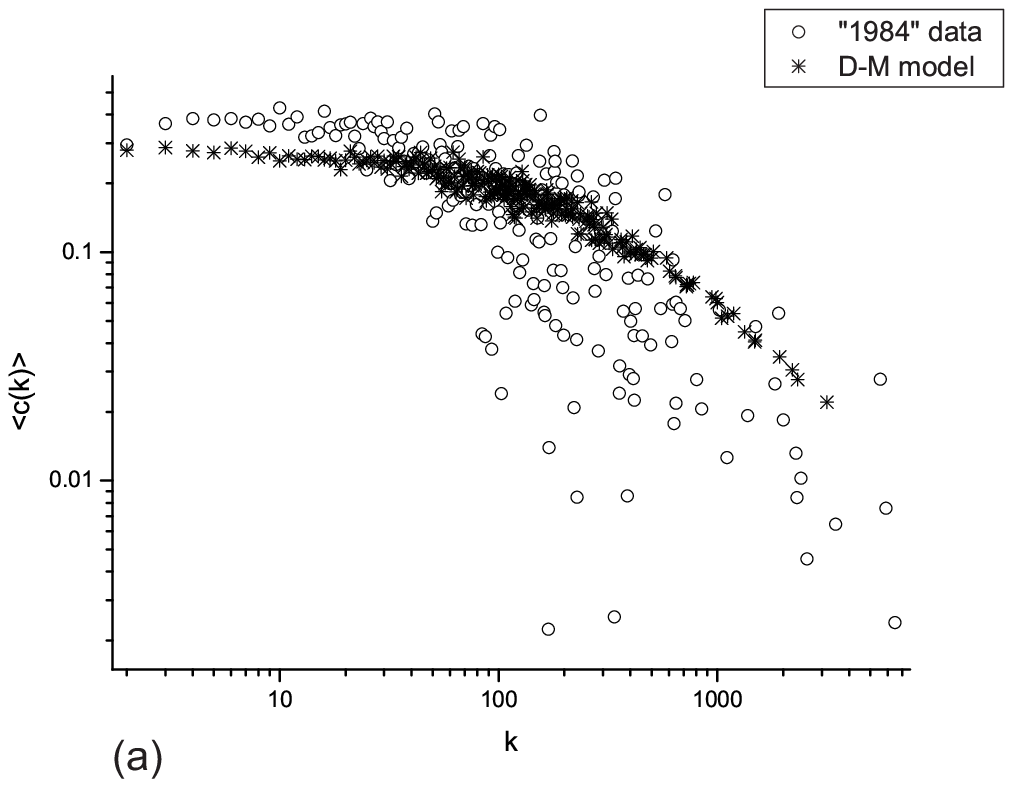}
         \includegraphics[width=0.48\textwidth]{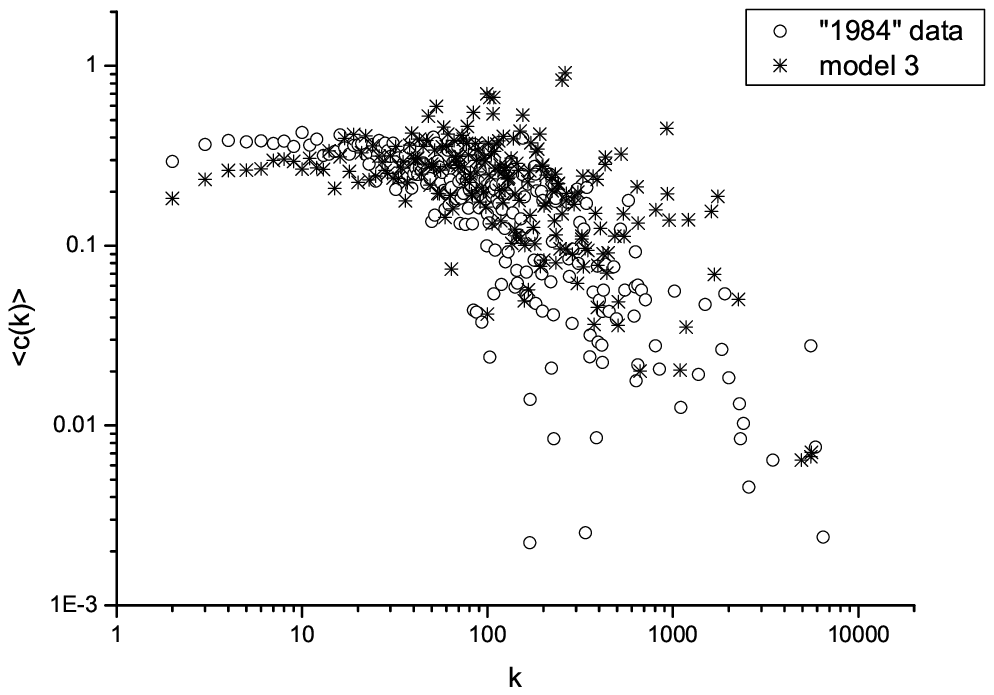}
         \caption{\label{f8} Average clustering coefficient versus the degree of the vertices. (a) Comparison between
         \textit{D-M model} and real data. (b) Comparison between \textit{model 3} and real data.}
    \end{figure}

The D-M model, model 3 and \textit{1984} data are compared in
Fig.\ref{f7} and Fig.\ref{f8}.  Zipf's analysis, in
Fig.\ref{f7}a, reveals that model 3 is better able than the D-M
model to describe the hierarchical behavior of the real data. In
 Fig.\ref{f8}b the measured average clustering
coefficient for model 3 results in a spread distribution. This is
evidence of a role differentiation in the vertices during the
evolution of the network, that is a self-organization of the
vertices in the different local structures. Nevertheless the main
global statistical properties of the network, as shown in
Fig.\ref{f7}b, are preserved.

\section{Conclusions}
In this work we analysed in detail the topology of human written
language, through a network representation of Orwell's
\textit{1984}. We performed average clustering coefficient and
nearest neighbors analysis, finding that two different vertex
behaviors clearly emerge. We performed entropic analysis that
allowed us to distinguish different roles of words. We studied the
relevance of second order correlations between vertices, finding
that those are essential properties of the network architecture.

We proposed a model for matching the identified empirical
behavior. This model included new growth mechanisms; a local
preferential attachment and the allocation of a set of preselected
vertices that have a structural rather than a functional purpose.

The degree of complexity of our model is greater than that of
classical models, but we can give a simple explanation for the
necessity to deal with such a degree of complexity. First of all,
classical scale free network models often don't reflect the
emergence of  Zipf's Law, that is they don't consider  the main
hierarchical architecture of the network.
 The random attachment to a set of preselected vertices, representing the main punctuation,   gives a solid structure to the network,
between which the scale free system can grow and allows the rhythm
of the novel to  emerge from their punctuation.

 We would like to
stress that nearest neighbors analysis for our network display
peculiar behavior, much more than those showed in this brief
review, and should be considered as the basis for the
understanding of network theory with a mixed local-global growth
mechanism. These considerations are relevant for all  natural
systems showing syntactic-like organization rules, that are
selective rules creating an intelligent ensemble from simple
elements.

Further empirical and theoretical research would be useful. Human
language is very important for the general study of network theory
because of its great availability, the precision of the data and
because we have a detailed knowledge of its local organizational
rules.
\begin{acknowledgments}
This research is part of the NET-ACE project, supported by the EC.
We would like to thank Pierpaolo Vivo for discussions and useful
suggestions.
\end{acknowledgments}


\end{document}